\documentclass[aps,prb,reprint,twocolumn,superscriptaddress,showpacs]{revtex4-1}
\usepackage{graphicx}
\usepackage{mathrsfs}
\usepackage{bm}
\usepackage{amsmath}
\usepackage{dcolumn}
\usepackage{epstopdf}
\usepackage{dsfont}
\usepackage{amssymb}
\usepackage{tabularx}
\usepackage{array}
\usepackage{float}
\usepackage{color}
\usepackage{epstopdf}
\usepackage{mathrsfs}
\usepackage[colorlinks, linkcolor=blue,anchorcolor=blue,citecolor=blue,urlcolor=blue]{hyperref}

\begin{document}

\title{Pure spin current and perfect valley filter  by designed  separation of the chiral states in two-dimensional honeycomb lattices} %
\author{Da-Ping Liu}
\affiliation{Department of Physics, Renmin University of China, Beijing 100872, China}
\author{Zhi-Ming Yu}
\email{zmyu@bit.edu.cn}
\affiliation{Beijing Key Laboratory of Nanophotonics and Ultrafine Optoelectronic Systems, School of Physics, Beijing Institute of Technology, Beijing 100081, China}
\affiliation{Research Laboratory for Quantum Materials, Singapore University of Technology and Design, Singapore 487372, Singapore}
\author{Yu-Liang Liu}
\email{ylliu@ruc.edu.cn}
\affiliation{Department of Physics, Renmin University of China, Beijing 100872, China}

\date{\today}

\begin{abstract}
We propose a  realization of pure spin  currents  and  perfect valley filters based on a quantum anomalous Hall  insulator, around which edge states with  up-spin and down-spin circulate.
By applying  staggered sublattice potential on the   strips along the   edges  of  sample, the edge states with down spin can be  pushed into the inner boundaries of the  strips while  the other edge states with up  spin  remain  on the outer boundaries, resulting in spatially separated chiral states with perfect  spin polarization.
Moreover, a  valley filter, which is immune to  short-range and smooth long-range scatterers, can be engineered by additionally applying boundary potentials  on the  outmost lattices of the sample.
We also find that the boundary potential can be used  to  control  the size effect  induced oscillation of  the inner chiral states.
The connection of the  boundary potential   to size effect  is revealed.
\end{abstract}

\pacs{73.43.-f, 72.25.-b, 73.63.-b, 71.70.Ej}

\maketitle
\section{Introduction}
The extensive research on spintronics and valleytronics unveil that
many binary degrees of freedom of electrons, e.g.,
spin and valley  can be used to improve the efficiency of electronic devices\cite{Hanson rmp 2007,Rycerz np 2007,xiao prl 2007,xiao prl 2012,xiao prl 2012}.
Economic consideration makes designing electronic devices with both
        various degrees of freedom and low power consumption quite an attractive study area.
Although various kinds of non-dissipative chiral (helical) modes have been theoretically proposed and experimentally realized\cite{Ezawa 2013,Ju 2015,Pan2 2015,Zhang 2011,Qiao 2011,Xiao 2011,Pan 2015,Rachel 2014},
        it is yet desirable  to present an easy way to spatially separate all the chiral states to obtain currents with a pure degree  polarization, such as    pure spin (valley)  current\cite{Yang2015,Jiang2015,li prl 2013}, as well as to  independently  manipulate   them.
Particularly, the chiral states pushed away from the boundaries will be
immune to the boundary perturbations\cite{Wang2016}, such as  the dangling bonds and  boundary defects, which generally are  obstacles for the application of the topological edge states.

 When a system   hosts    two or more spatially separated chiral  states, the influence of  finite size effect on the chiral states may be significant\cite{Zhou 2008,Ezawa2 2013,Fu2014}.
The size effect can induce a undesired  gap  if the two chiral states with  same momentum  have overlap in the real space.
Generally, one can eliminate the gap by shifting the momentum of the  two chiral states\cite{Zhou 2008}.
However, the influence of size effect on the spatial distribution of the  chiral states would persist even if the   spectrum is gapless.

In this paper,  we propose a scheme for  spatially separating all the  edge states of a quantum anomalous Hall (QAH) insulator with Chern number $|{\cal C}|=2$ based on two-dimensional honeycomb lattices to realize pure spin currents.
The QAH effect can be introduced  into honeycomb lattices if one considers the  Haldane term, which was artificially proposed at first\cite{Haldane 1988}.
But  later, it turns out that the Haldane term can be realized in  laser-driven systems \cite{Kitagawa2011,Scholz2013,Ezawa 2013}, where the low-energy electrons submit to the Floquet theory\cite{Kitagawa2011,Cayssol2013,Lopezl2015,Wang2013,Rechtsman2013}.
Usually, the honeycomb lattices also may possess sizeable   intrinsic spin-orbit coupling (SOC)\cite{Zhao2016,Liu 2011,Song 2014},
thus in such case  the appearance of QAH effect requires that  the Haldane term should dominate the  SOC.

In the presence of Haldane term and SOC, a QAH insulator with $|\mathcal{C}|= 2$ can be generated.
At each edge of the QAH sample, the two unipropagating edge states are of up-spin and down-spin respectively and thus, the edge current is spin unpolarized.
By applying staggered sublattice potential  to narrow strips along the two edges, one pair of the edge states with down-spin can be  pushed to the inner boundaries of the potential  regions, while the other with up-spin will be  still at the outer boundaries.
With the help   of  local terminals,  the  spatially separated pure spin currents  can be  independently  extracted out.
Moreover, by additionally applying potentials  on the  outmost lattices of the sample, all the chiral states will reside  around the valley points and  the velocity
of them can be   valley dependent\cite{Yao2009}, giving rise to a perfect valley filter, as the counterpropagating chiral  states are well separated in both real and momentum space.
Also, the  spin and valley index of the inner chiral  states  can be switched by changing the sign of  Haldane term (e.g.
 switching the polarized direction of the  circularly polarized laser) and the direction of the staggered potential.
The finite  size effect will induce an oscillation on the spatial distribution of the inner chiral states.
Interestingly, we find that a local rather than global manipulation tuning the boundary potential can effectively control the size effect induced oscillation.

Our paper is organized as follows. In Sec. \ref{Model},
we introduce a model Hamiltonian of honeycomb lattices with intrinsic  SOC and  Haldane term under a spatial dependent  potential.
Section \ref{curr} describes the spatial separation of  chiral edge states and the realization of  perfect valley filter.
We investigate the size effect  induced oscillation of the inner chiral states in   Sec. \ref{size}.
 Finally, we present some discussions and  conclusions  in Sec. \ref{dis}.

\section{Model Hamiltonian} \label{Model}
We start from a tight-binding (TB) Hamiltonian defined on a two-dimensional honeycomb lattice $\mathcal{H}=\mathcal{H}_0+\mathcal{H}_{1}$ with
\begin{eqnarray}
{\cal H}_{0} & = & -t\sum_{\langle i,j\rangle\alpha}c_{i\alpha}^{\dagger}c_{j\alpha}+i\frac{\lambda}{3\sqrt{3}}\sum_{\langle\langle i,j\rangle\rangle\alpha\beta}v_{ij}c_{i\alpha}^{\dagger}\tau_{\alpha\beta}^{z}c_{j\beta}\nonumber \\
 &  & +i\frac{\lambda_{\omega}}{3\sqrt{3}}\sum_{\langle\langle i,j\rangle\rangle\alpha}v_{ij}c_{i\alpha}^{\dagger}c_{j\alpha},\label{eq:Ham0}
\end{eqnarray}
where $\tau^{z}$ is  $z$-component  Pauli matrix.
$t$ is the nearest-neighbor hopping.
$\lambda$ and  $\lambda_{\omega}$ represent the strength of  intrinsic SOC and  Haldane term respectively.
The  Haldane term can be induced by an off-resonance coherent and circularly polarized laser beam.
$v_{ij}=+(-)$ if  the hopping from site $j$ to site $i$ is anticlockwise (clockwise) with respect to the positive $z$ axis\cite{Kane2005}.
And
\begin{eqnarray}
{\cal H}_{1} & = \sum_{i}\mu_{i}\Delta c_{i\alpha}^{\dagger}c_{i\alpha},\label{eq:Ham1}
\end{eqnarray}
represents a  staggered sublattice potential  with $\mu_{i}=+(-)$ for the $A$ ($B$) site, which  may  be induced by substrate\cite{Zhou2007} or by simply applying an external electric field if the system is  buckled\cite{Yu2015,Liu22011}.

To gain a transparent  understanding of the topological properties of $\mathcal{H}$
and the following discussions of spatial separation of the chiral edge modes, we write down the low-energy Hamiltonian  around the valley points,
\begin{eqnarray}
H & = & v_{\rm F}\left(\eta k_{x}\sigma_{x}+k_{y}\sigma_{y}\right)+\left[\Delta-\eta\left(\lambda s_{z}+\lambda_{\omega}\right) \right]\sigma_{z},\label{eq: Lham}
\end{eqnarray}
where $v_{\rm F}=\sqrt{3}a_0t/2$ is the Fermi velocity with  $a_0$  the lattice constant.
Pauli matrices $\boldsymbol{\sigma}$ act on the sublattice pseudospin,  $\eta=+$ ($-$) denotes the  valley index $K$ ($K'$), and $s_{z}=+$ ($-$) represents up (down) spin.
According to Hamiltonian (\ref{eq: Lham}),  the valley-contrasting topological charge is\cite{Pan2 2015,Martin2008,Yao2009}
\begin{eqnarray}
{\cal C}_{s_{z},\eta} & = & \frac{\eta}{2}\times\text{sgn}[\Delta-\eta(\lambda s_{z}+\lambda_{\omega})].
\end{eqnarray}
In the absence of potential  $\Delta=0$, the system is a QAH (quantum spin Hall) insulator  if $|\lambda_{\omega}|>|\lambda|$ ($|\lambda_{\omega}|<|\lambda|$).
Since we are interested in a QAH sample here,   without loss of generality, we let $\lambda_{\omega}>\lambda>0$, and the following discussions can be directly extended  to the case of $\lambda_{\omega}<0$ and $|\lambda_{\omega}|>|\lambda|$.
If $|\Delta|<\lambda_{\omega}-\lambda$,
        the topological properties of system are the same with that of $\Delta=0$.
On the other hand, if $|\Delta|>\lambda_{\omega}+\lambda$,
        then ${\cal C}_{s_{z},\eta}(\Delta)=\frac{\eta}{2}\text{sgn}[\Delta]$
        is only determined by the sign of $\Delta$.
The interesting case is that  when $|\Delta|\in(\lambda_{\omega}-\lambda,\lambda_{\omega}+\lambda)$,
${\cal C}_{+,\eta}(\Delta)=\frac{1}{2}$
and ${\cal C}_{-,\eta}(\Delta)=\frac{\eta}{2}\text{sgn}[\Delta]$.
In this case, band structure of electron with $s_{z}=1$ is still  topology nontrivial
        while that of electron with $s_{z}=-1$ is topology trivial,
        as the total Chern number of up (down) spin is $\sum_\eta {\cal C}_{+,\eta}=1$ ($\sum_\eta {\cal C}_{-,\eta}=0$).
Thus, the  potential $|\Delta|\in(\lambda_{\omega}-\lambda, \lambda_{\omega}+\lambda)$ only changes the topological properties of down-spin.

\begin{figure}[t]
\includegraphics[width=8cm]{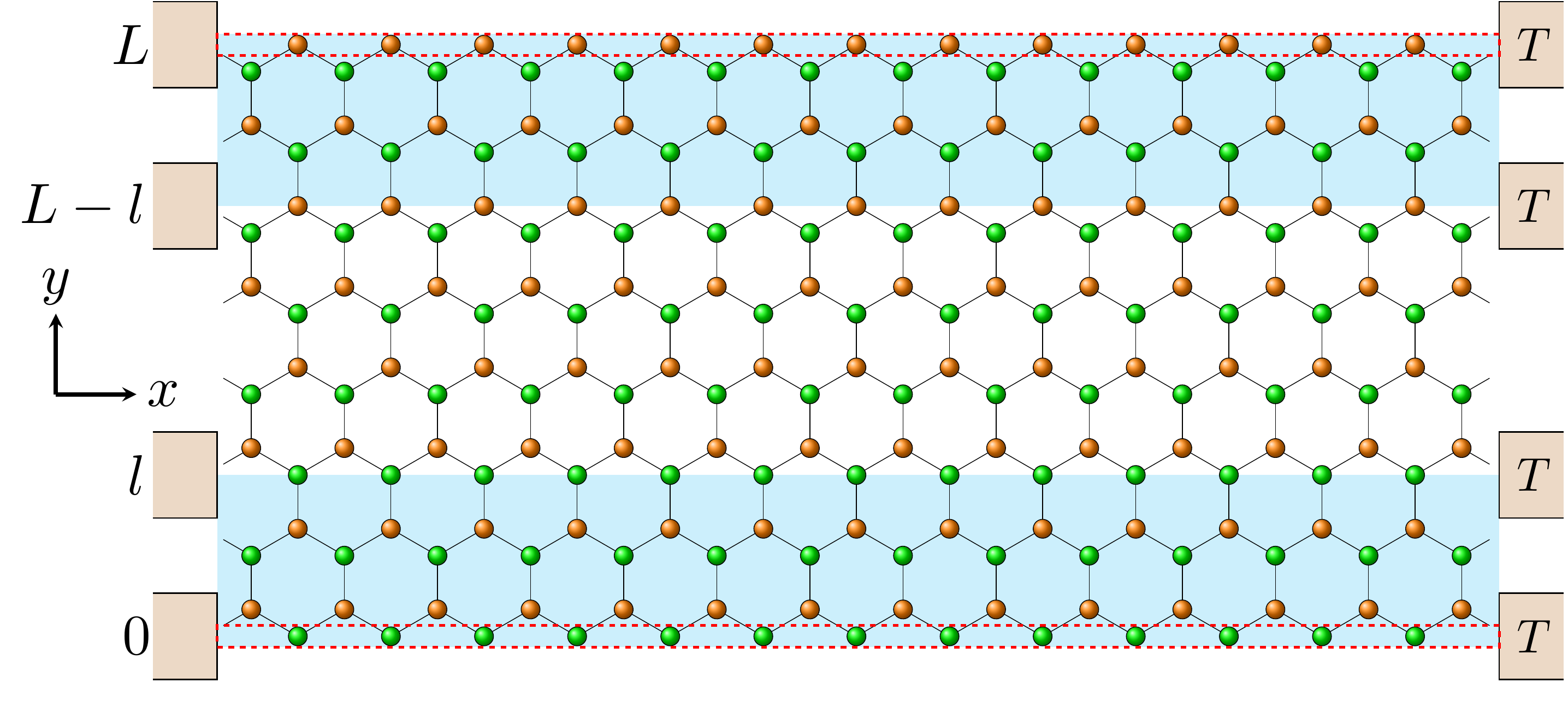}

\protect\caption{(Color online) Schematic of a QAH sample with two long strips geometry. The two sublattices $A$ and $B$ are denoted by olive and red balls respectively. And the profile of the $y$-dependent  staggered sublattice potential  is shown by the color in the nanoribbon. The  local  terminal ($T$) can extract the spin perfectly
polarized current out.
} \label{Fig. 1}
\end{figure}

Based on  the above discussions,
        we consider a QAH sample with strips applied on  different  potential, as shown in Fig. \ref{Fig. 1}.
The potential $\Delta$ is chosen  to be nonzero in region I ($0<y<l$) and region III ($L-l<y<L$),
        and vanishing in region II ($l<y<L-l$).
One can find  that if $|\Delta_{\rm I(III)}|\in(\lambda_{\omega}-\lambda,\lambda_{\omega}+\lambda)$,
then for the electrons with $s_{z}=-1$, region II will be  topologically distinct from the others, while for the electrons with $s_{z}=1$,  the three regions  are still topologically equal.
According to the bulk-boundary correspondence\cite{Volovik2003}, there would be chiral modes with $s_{z}=-1$ appear at the boundaries of region II.
While  the pair of edge states with $s_{z}=1$ would be still at the outer boundaries.
Moveover, from the index theorem\cite{Martin2008,Semenoff2008,Volovik2003}, one has $v_{s_{z},\eta}^{\rm I(III)}  =  {\cal C}_{s_{z},\eta}(\Delta_{\rm I(III)})-{\cal C}_{s_{z},\eta}(\Delta_{\rm II})$, where $v_{s_{z},\eta}^{\rm I(III)}$ is the number  of chiral modes with index $s_z$ and $\eta$  located at the interface of region I (III) and region II.
In the case of $|\Delta_{\rm I(III)}|\in(\lambda_{\omega}-\lambda,\lambda_{\omega}+\lambda)$, one has
\begin{eqnarray}
v_{+,\eta}^{\rm I(III)}=0 & ;\  & v_{-,\eta}^{\rm I(III)}=\frac{1}{2}\left[\eta\times\text{sgn}\left(\Delta_{\rm I(III)}\right)+1\right],\label{eq: v}
\end{eqnarray}
indicating  that the inner chiral states must be of  $s_{z}=-1$ and of $\eta=\text{sgn}[\Delta_{\rm I(III)}]$.
And the valley index of the inner chiral state  can be switched  by changing the sign of $\Delta_{\rm I(III)}$.

\begin{figure*}[htbp]
\includegraphics[width=18.cm]{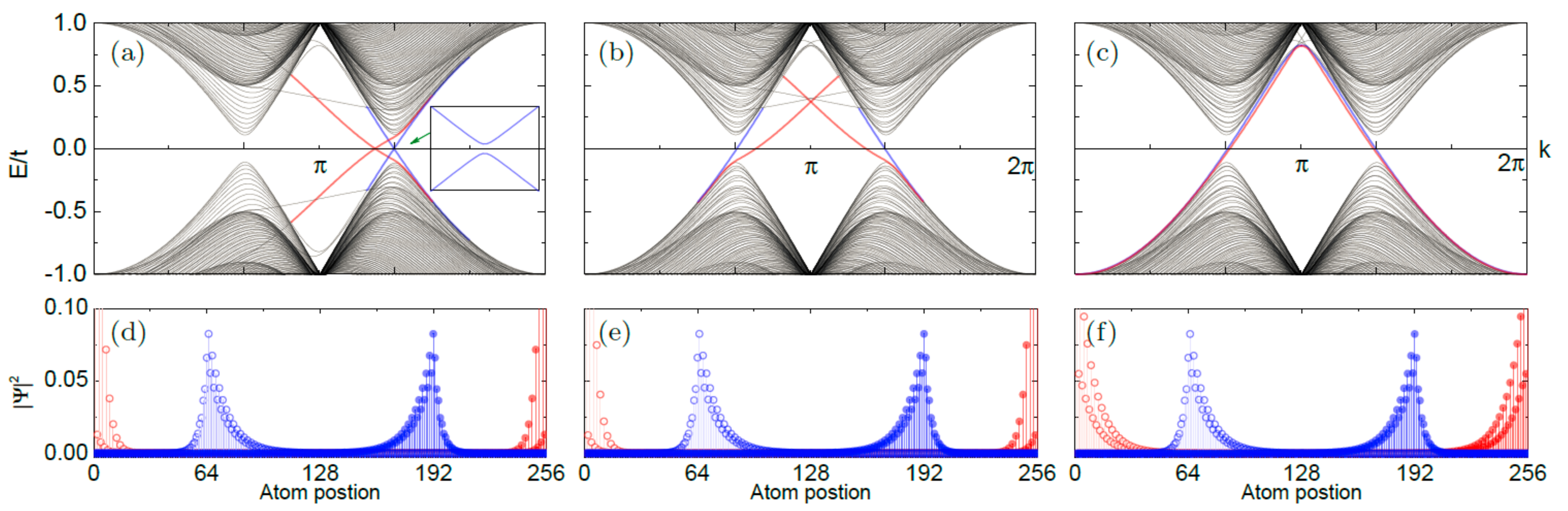}

\protect\caption{(Color online)
Band structures of the QAH nanoribbon for (a)  $\Delta_{\rm I}=\Delta_{\rm III}=0.4\,t$, (b)  $\Delta_{\rm I}=-\Delta_{\rm III}=0.4\,t$  and (c)  $\Delta_{\rm I}=-\Delta_{\rm III}=0.4\,t$ with boundary potential  $\Delta_{\rm I}^{\prime}=-\Delta_{\rm III}^{\prime}=t$.
(d-f) Corresponding spatial distribution of the four chiral states near Fermi energy.
The chiral states with up-spin (down-spin) are denoted by red (blue) lines.
And  the solid (hollow) circles correspond to right- (left-)propagating chiral states.
In the figures, we choose $\lambda=0.2\,t$, $\lambda_{\omega}=0.3\,t$, $L=256$ atoms, and $l=64$ atoms.
 } \label{Fig. 2}
\end{figure*}

\section{Pure spin  current  and perfect valley filter} \label{curr}
In the present model, spin is always a good quantum number.
For the pristine QAH sample ($\Delta_{\rm I}=\Delta_{\rm III}=0$),  the edge currents propagating along the  edges of sample  are spin unpolarized, as they simultaneously contain up-spin and down-spin.
By applying suitable potential $|\Delta_{\rm I(III)}|\in(\lambda_{\omega}-\lambda,\lambda_{\omega}+\lambda)$,
four spin perfect  polarized currents are generated  propagating   along  the four interfaces $y=0,~l,~L-l,~\text{and}~L$ respectively
[Fig. \ref{Fig. 2}(a) and \ref{Fig. 2}(b)].
With the  help of  local terminals (schematically shown in Fig. \ref{Fig. 1}), pure spin currents can be extracted out whenever  the distance between the four interfaces (e.g. $l$ and $L-2l$) are much larger than the  spatial broadening of  the chiral states, which generally can be satisfied by choosing  suitable parameters as shown in Fig. \ref{Fig. 2}.
Also,  the valley index of the inner chiral states  can be  independently manipulated
        by switching the sign of $\Delta_{\rm I}$ or  $\Delta_{\rm III}$ [Eq. (\ref{eq: v})].
Hence the two inner chiral states can have same ($\Delta_{\rm I}\Delta_{\rm III}>0$) or different  ($\Delta_{\rm I}\Delta_{\rm III}<0$) valley index.
Particularly, when $\Delta_{\rm I}\Delta_{\rm III}<0$,
the two inner chiral states  locate at different valley and would be always gapless regardless of width of region II [Fig. \ref{Fig. 2}(b)].
Generally, the applied electric field and substrate can induce a weak  Rashba SOC\cite{Kane2005}, which couples different spin modes\cite{Sengupta2006}. However, the pure spin currents proposed here are robust against a weak Rashba  SOC, as  all the chiral (spin) modes are spatially separated.

In the present model, a perfect valley filter also can be engineered  by additionally applying boundary   potentials $\Delta_{\rm I(III)}^{\prime}$  on  the outermost sites of sample (marked by the red frame in Fig. \ref{Fig. 1}).
Let us start from Fig. \ref{Fig. 2}(b) where $\Delta_{\rm I}\Delta_{\rm III}<0$.
The outer edge states (denoted as red lines), which  do not have well defined valley index in such case,
        mainly concentrate on the boundaries of sample as shown in Fig. \ref{Fig. 2}(e).
By tuning the boundary potentials $\Delta_{\rm I(III)}^{\prime}$,
        the two outer edge band dispersion can  bend upward and  gradually merge into the bulk conduction band\cite{Yao2009}.
At the same time, the zero mode of the outer edge states would  reside around   the corresponding valley points [Fig. \ref{Fig. 2}(c)].
Consequently, all the four chiral states possess well-defined valley index and the  velocity of them is valley dependent.
Since  the counter-propagating chiral states are well separated in both real and momentum space,
the valley filter and pure spin currents proposed here  would be immune to  short-range  scatterers, as well as the long-range scatterers that are
smooth at the atomic scale.

\section{size effect} \label{size}
In this section, we study the finite size effect on the chiral modes.
Without resorting  to the detailed analysis, one can obtain several key features of the size effect: (1)  Size effect  is induced by  the  interfaces of system, including the boundaries and the inner interfaces. The boundaries of sample would have stronger limit effect than that of the inner interface, as the electrons can pass through the inner interfaces but are forbidden to be out of sample.
(2) If the chiral modes intersect in the momentum space, the  size effect can produce a width dependent gap [Fig. \ref{Fig. 2}(a)].
(3) The wavefunction of the chiral mode is determined by the boundary conditions.  Hence the inner chiral states in the present model  would have   distinct wavefunction  compared with the chiral modes in a perfect domain wall and the edge states located at the boundary of sample.
The size effect on the  spectrum of chiral modes  have been studied in previous wroks\cite{Ezawa2 2013,Zhou 2008}.
In this paper, we focus on the influence of  size effect on the wavefunction of the inner chiral modes.
We find that the inner chiral mode would  possess peculiar oscillated  wavefunction induced by the size effect  and interestingly, this  oscillation can be controlled by a local manipulation tuning the boundary potential.

\begin{figure}[t]
\includegraphics[width=8.cm]{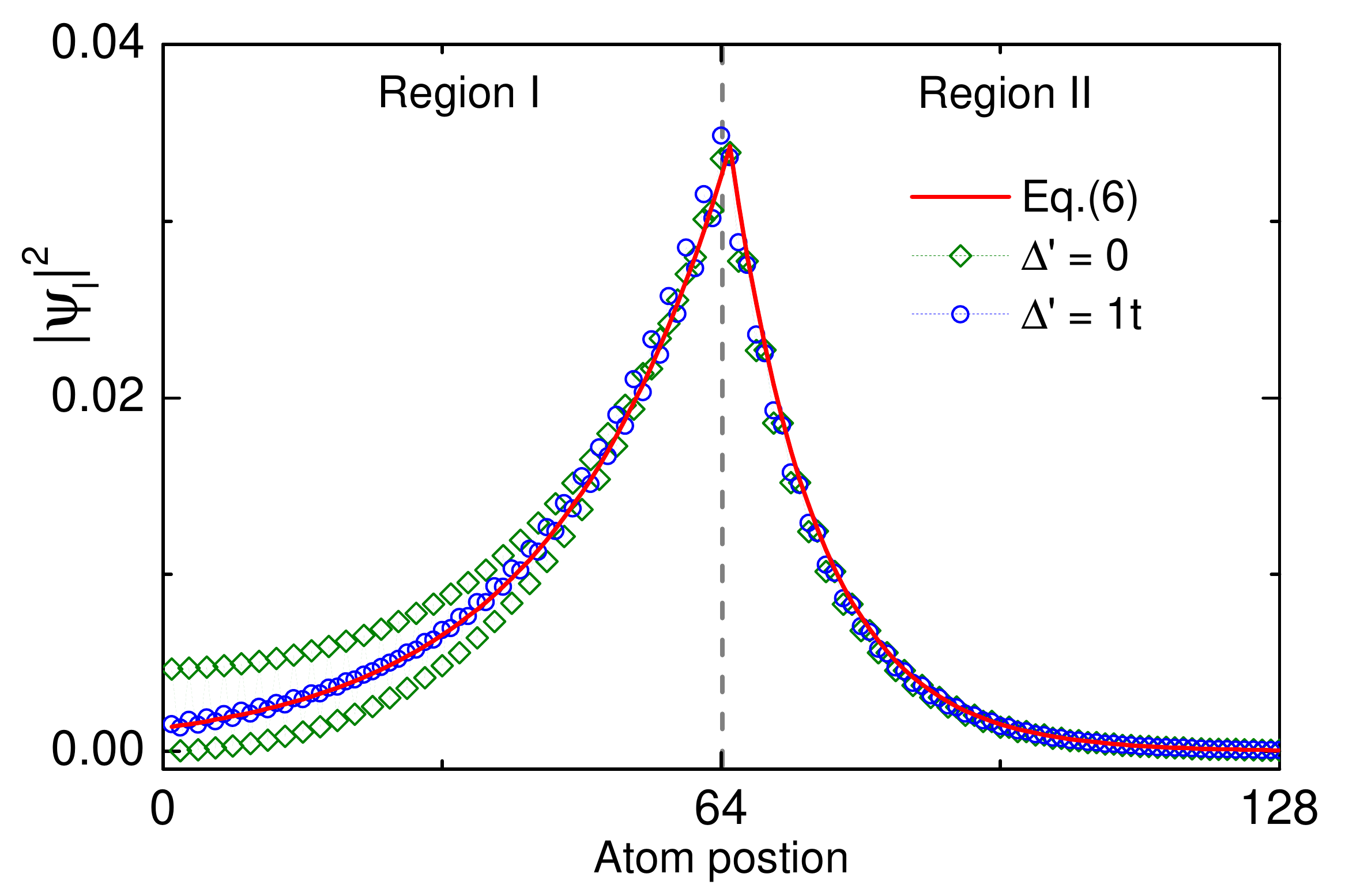}
\protect\caption{(Color online)The influence of  boundary potential $\Delta^{\prime}$ on the size effect induced oscillation. Spatial distribution of the inner chiral  state $|\Psi_l(x,y)|^2$ at $K$ valley point for different $\Delta^{\prime}$ are plotted.
The red solid lines are from Eq. (\ref{eq:wf}) and the hollow circles are solutions from the TB model with different  $\Delta^{\prime}$. In the figure, we choose $\lambda=0.2\,t$, $\lambda_{\omega}=0.3\, t$, $\Delta_{\rm I}=0.15\,t$, $\Delta_{\rm III}=-0.15\,t$, $l=64$ atoms and the ribbon width $L=256$ atoms.
} \label{Fig. 3}
\end{figure}

The inner chiral mode is reminiscent of the chiral mode in a domain wall   respecting Jackiw-Rebbi solution\cite{Jackiw1976}, of which the general form is\cite{Pan 2015}:
\begin{eqnarray}
\Psi_{D}(x,y) & =& \left(\begin{array}{c}
\phi_{A}(y)\\
\phi_{B}(y)
\end{array}\right)= \frac{1}{{\cal N}}\left(\begin{array}{c}
1\\
-i
\end{array}\right)e^{ik_{x}x+(y-l)\kappa(y)},\label{eq:wf}
\end{eqnarray}
where the location of domain wall is chosen to be $y=l$, ${\cal N}$ is the normalization constant, and $\phi_{A(B)}$ is the wavefunction of $A$ ($B$) sublattice.
When ribbon width is infinite that $L\gg l\gg 0$,  there is no size effect and the inner chiral  state $\Psi_{l}(x,y)$ (located at $y=l$) would be the same as $\Psi_{D}(x,y)$, with $\kappa(y)=\kappa_{\rm I}\equiv(\Delta_{\rm I}+\lambda-\lambda_{\omega})/v_{\rm F}$ for $y<l$ and $\kappa(y)=\kappa_{\rm II}\equiv(\lambda-\lambda_{\omega})/v_{\rm F}$ for  $y>l$ (here we choose  $\Delta_{\rm I}>0$, and $\Delta_{\rm I}\Delta_{\rm III}<0$).
When $L$ is finite but $\kappa_{\rm I} l\gg1$,  $\Psi_{l}(x,y)$  would be  still almost the  same as  $\Psi_{D}(x,y)$.
On the other hand, when    $\kappa_{\rm I}l\simeq 1$, $\Psi_{l}(x,y)$ would  deviate from $\Psi_{D}(x,y)$, corresponding to the presence of  a   finite size effect.
In Fig. \ref{Fig. 3}, we plot the spatial distribution  of the inner chiral  state $|\Psi_{l} (y)|^2$ with $k_y=2\pi/3$ ($K$ valley point) and the potential $\Delta_{\rm I}$   chosen to satisfy  $\kappa_{\rm I}l\simeq 1$.
In  region I, $|\Psi_{l} (y)|^2$ (olive rhombus) presents an oscillation  around  $|\Psi_{D}(y)|^2$ (red line).
One can understand this oscillation by the geometry of the zigzag  nanoribbon.
For a zigzag  nanoribbon, $\phi_A(y)$ and $\phi_B(y)$ obey symmetric quadratic differential equation but different  boundary conditions \cite{Brey2006} that $\phi_B(0)=0$ while $\phi_A(0)$  is finite except in the limit of no size effect that  $\kappa_{\rm I}l\gg1$,
thus the behavior of $\phi_A(y)$ and $\phi_B(y)$ are  very different when   $\kappa_{\rm I}l\simeq 1$, leading to the  oscillation.

The oscillation amplitude of  $|\Psi_{l} (y)|^2$ generally can be controlled by changing    $l$ and $\kappa_{\rm I}$.
However, we find that a local manipulation tuning the boundary potential   also can be used to control the size effect induced oscillation.
With finite $l$, one does not have $|\phi_A(0)|=|\phi_B(0)|$ in a zigzag  nanoribbon, however $|\phi_A(0^+)|=|\phi_B(0^+)|$ is possible, in such case the behavior of  $|\phi_A(y)|$ and $|\phi_B(y)|$  in region I will be identical except $y=0$ and thus the  size effect induced oscillation is eliminated.
To realize  above situation, one needs to introduce  a delta function [$\delta(0)\sigma_z$] to make  $\phi_{A(B)}(y)$  to be discontinuous at $y=0$.
The delta function $\delta(0)\sigma_z$ in low-energy continuum model corresponds to an abrupt change of  potential at boundary $y=0$ in a TB lattice  model, which is nothing but tuning the boundary potential $\Delta^{\prime}$.
The  influence of $\Delta^{\prime}$  on the oscillation is   shown in Fig. \ref{Fig. 3} too.
By increasing $\Delta^{\prime}$ from zero, $|\phi_B(0^+)|$ will increase from zero, which  inevitably leads to the decrease of $|\phi_A(0^+)|$ as the total wavefunction $\Psi_l(y)$ is normalized.
 When $\Delta^{\prime}$ approaches a critical value $\Delta^{\prime}_c$, one would  have $|\phi_B(0^+)|=|\phi_A(0^+)|$ and the oscillation is  eliminated.
If one keeps increasing  $\Delta^{\prime}$,  $|\phi_B(0^+)|$  will be  larger than $|\phi_A(0^+)|$ and the oscillation will  reappear.

The connection of the boundary potential to size effect induced oscillation can be quantitatively understood within a Dirac equation.
Consider the  TB Hamiltonian with boundary potential $\Delta^{\prime}$,
\begin{eqnarray}
{\cal H}^{\prime}&=&{\cal H}+\Delta^{\prime}c_{y=0,A}^{\dagger}c_{y=0,A},
\end{eqnarray}
with  ${\cal H}={\cal H}_0+{\cal H}_1$.
The wavefunction of the chiral state $\Psi _l(y)$ can be obtained by  the corresponding  Dirac equation\cite{Brey2006,Wang2014,Nakada1996} defined around $K$ valley ($\Delta_{\rm I}>0$)
\begin{eqnarray}
H' & = & v_{F}\left(k_{x}\sigma_{x}-i\partial_{y}\sigma_{y}\right)+\left(\bar{\Delta}(y)+\bar{y}\Delta^{\prime}\delta_{y=0}\right)\sigma_{z},
\label{hp}
\end{eqnarray}
where $\bar{\Delta}(y)=\Delta_{\rm I}+\lambda-\lambda_{\omega}$ ($=\lambda-\lambda_{\omega}$) in region $y<l$ ($y>l$) and $\bar{y}=\sqrt{3}a_0/2$ is the average distance of site $A$ (or $B$) along the $y$ direction.
Integrating Eq. (\ref{hp}) around $y=0$ and considering that $\phi_B(0)=0$, one has
\begin{eqnarray}
\Delta^{\prime}\bar{y}\phi_{A}(0) & = & \Delta^{\prime}\bar{y}\phi_{A}(0^{+})=iv_{F}\phi_{B}(0^{+}),
\end{eqnarray}
which directly show that when $\Delta^{\prime}=t$, $|\phi_A(0^+)|=|\phi_B(0^+)|$.
From Fig. \ref{Fig. 2}, one also observes that the energy of the inner chiral states located at valley point is close to zero $\varepsilon_l(k_y=2\pi/3)\approx  0$.
In the case of $\bar{\Delta}(y<l)\gg \varepsilon_l(2\pi/3)$, one can approximately let  $\varepsilon_l(2\pi/3)=0$ and then has
\begin{eqnarray}
\Psi_{l}(x,y) & \approx & \frac{1}{{\cal N}_{l}}\left(\begin{array}{c}
e^{y\kappa_{\rm I}}+\gamma e^{-y\kappa_{\rm I}}\\
-i\left(e^{y\kappa_{\rm I}}-\gamma e^{-y\kappa_{\rm I}}\right)
\end{array}\right),\label{eq: wl}
\end{eqnarray}
for $0<y<l$, where ${\cal N}_{l}$ is the normalization constant and
\begin{eqnarray}
\gamma & = & \frac{v_{\rm F}-\bar{y}\Delta^{\prime}}{v_{\rm F}+\bar{y}\Delta^{\prime}}
=\frac{t-\Delta^{\prime}}{t+\Delta^{\prime}}.
\end{eqnarray}
For $y>l$, $\Psi_{l}(x,y)$ would have  the same expression with $\Psi_{D}(x,y)$ up to a normalization constant whenever $L-2l\gg v_F/\bar{\Delta}(y>l)$.
From Eq. (\ref{eq: wl}), one knows that  when $\Delta^{\prime}=t$,    $\gamma=0$ and   $\Psi_l(x,y)$ will be  identical  with $\Psi_{D}(x,y)$ in both regions $0<y<l$  and  $y>l$ (up to a normalization constant), and thus the size effect induced oscillation would be approximately eliminated, as shown in Fig. \ref{Fig. 3}.
When $\Delta^{\prime}=t$, $\Psi_l(x,y)$ has a rather simple expression and then  the spatial distribution of the inner chiral modes can be directly obtained,
  which is helpful to the precise manipulation of these  modes.

\section{Discussion and conclusion }\label{dis}
We have proposed a model to realize  pure spin  current and perfect valley filter by spatially separating all
the chiral  modes of a QAH sample.
Also we have presented a local control of the finite size effect induced oscillation of the inner chiral  states.

Many of the already known  two-dimensional honeycomb lattices, especially the buckled materials,  may be used to realize the present  proposals.
First, the   staggered sublattice potential can be easily induced and  tuned by a perpendicular electric field in buckled systems.
Second, buckled systems generally  possess sizeable intrinsic SOC\cite{Liu 2011}.
Third, the  strength of Haldane term can be tuned by changing the  intensity and frequency of the circularly polarized laser beam and the situation of $\lambda_{\omega}>\lambda$ may be  satisfied\cite{Ezawa 2013,Dahlhaus2015}.
Thus, silicene, germanene and stanene are suitable candidates\cite{Liu22011},
especially  that silicene field-effect transistors have been recently fabricated at room
temperature\cite{Tao} and germanene has been successfully synthesized on a  band gap material\cite{Zhang}.
Notice that to realize the present proposal in a highly integrated microelectronic device is a challenging task, but
our investigation undoubtedly  provides an  opportunity to generate the pure spin currents and perfect valley filter.

Overall, our proposals of generating pure spin  currents and designing perfect valley filter   may be realized in many honeycomb lattices.
And the ease of electric field control of low-dimensional buckled materials would give rise to a wide application scope of the present proposal.

\begin{acknowledgments}
The authors acknowledge discussions with Shengyuan A. Yang and Yugui Yao.
This work is supported by the Fundamental Research Funds for the Central Universities, the Research Funds of Renmin University of China (14XNLQ03),
and the National Basic Research Program of China (Grant No. 2012CB921704).

\end{acknowledgments}

\end{document}